\documentclass[%
 reprint,
 amsmath,amssymb,
 aps,
]{revtex4-2}

\usepackage{color}
\usepackage{graphicx}
\usepackage{dcolumn}
\usepackage{bm}
\usepackage{algorithm}
\usepackage{algpseudocode}

\newcommand{\eref}[1]{Eq.~(\ref{#1})}
\newcommand{\fref}[1]{Fig.~\ref{#1}}

\newcommand{\tref}[1]{Table~\ref{#1}}
\newcommand{\Aref}[1]{Appendix~\ref{#1}}
\newcommand{\Alref}[1]{Algorithm~\ref{#1}}

\begin{document}

\preprint{123}

\title{Deep Domain Adversarial Adaptation for Photon-efficient Imaging}

\author{Yiwei Chen}
\affiliation{State Key Laboratory of Industrial Control Technology, Hangzhou, 310027, P. R. China}
\affiliation{College of Control Science and Engineering, Zhejiang University, Hangzhou, 310027, P. R. China}
\author{Gongxin Yao}
\affiliation{State Key Laboratory of Industrial Control Technology, Hangzhou, 310027, P. R. China}
\affiliation{College of Control Science and Engineering, Zhejiang University, Hangzhou, 310027, P. R. China}
\author{Yong Liu}
\affiliation{State Key Laboratory of Industrial Control Technology, Hangzhou, 310027, P. R. China}
\affiliation{College of Control Science and Engineering, Zhejiang University, Hangzhou, 310027, P. R. China}
\author{Hongye Su}
\affiliation{State Key Laboratory of Industrial Control Technology, Hangzhou, 310027, P. R. China}
\affiliation{College of Control Science and Engineering, Zhejiang University, Hangzhou, 310027, P. R. China}
\author{Xiaomin Hu}
\affiliation{CAS Key Laboratory of Quantum Information, University of Science and Technology of China, Hefei, 230026, P. R. China}
\affiliation{CAS Center For Excellence in Quantum Information and Quantum Physics, University of Science and Technology of China, Hefei, 230026, P. R. China}
 
\author{Yu Pan}
\email{ypan@zju.edu.cn}
\affiliation{State Key Laboratory of Industrial Control Technology, Hangzhou, 310027, P. R. China}
\affiliation{College of Control Science and Engineering, Zhejiang University, Hangzhou, 310027, P. R. China}

\date{\today}

\begin{abstract}
Photon-efficient imaging with the single-photon light detection and ranging (LiDAR) captures the three-dimensional (3D) structure of a scene by only a few detected signal photons per pixel. However, the existing computational methods for photon-efficient imaging are pre-tuned on a restricted scenario or trained on simulated datasets. When applied to realistic scenarios whose signal-to-background ratios (SBR) and other hardware-specific properties differ from those of the original task, the model performance often significantly deteriorates. In this paper, we present a domain adversarial adaptation design to alleviate this domain shift problem by exploiting unlabeled real-world data, with significant resource savings. This method demonstrates superior performance on simulated and real-world experiments using our home-built up-conversion single-photon imaging system, which provides an efficient approach to bypass the lack of ground-truth depth information in implementing computational imaging algorithms for realistic applications.
\end{abstract}

\maketitle

\section{Introduction}
Single-photon LiDAR systems have achieved dramatic success in 3D imaging over the past decades \cite{kirmani2014first,lussana2015enhanced, altmann2018quantum, musarra2019non, li2020single}. Benefiting from the high sensitivity of single-photon avalanche diode (SPAD) detector \cite{hadfield2009single, nicolich2019universal}, single-photon LiDAR provides better precision over a longer distance as compared to the traditional systems based on the photomultiplier tube \cite{dolgoshein2006status}. In practice, a photon counting histogram is collected at each scanned point with the time-correlated single photon counting (TCSPC) technique \cite{o2012time}, which records the timestamps of emitted and reflected photons for the time-of-flight (ToF) calculation. However, due to practical constraints on optical flux and integration time, the average number of detected photons per pixel, including signal and noise counts, could be less than a few dozens. This calls for photon-efficient imaging \cite{shin2015photon,shin2016photon}, whose goal is to recover a precise depth image from a sparsely distributed histogram of few photon counts.


Machine learning methods, including statistical \cite{shin2016photon, rapp2017few} and deep learning models \cite{lindell2018single, peng2020photon, zang2021non}, have been proposed for photon-efficient imaging. Particularly, deep learning models have demonstrated remarkable performance in terms of reconstruction accuracy and inference time. Deep learning models need huge amounts of data for training, whereas it is extremely difficult to acquire enough annotated real-world data. As a result, the existing models are trained on simulated datasets for which the SBR and other hardware-specific properties have to be assumed. For example, SBR could be affected by light condition, object distance or noise inherent in the hardware, such as the dark counts of SPAD and pump light \cite{rehain2020noise}. Hardware-specific properties also include the shape of a pulse \cite{weiner2000femtosecond}. Therefore, the model trained on simulated dataset may not generalize well to realistic hardware or environment for which SBR or pulse shape are different, causing the domain shift problem \cite{torralba2011unbiased, weiss2016survey,tang2022generalizable}. Although a calibration step can be conducted to estimate the realistic parameters, regenerating enough simulated data and retraining the model to fit the target domain requires tens of hours and a huge amount of computational resources, which is highly expensive and unscalable. Considering that ground-truth label is basically inaccessible in a real-world experiment, training the deep learning model with real-world data is even more impossible.  

Deep domain adversarial adaptation \cite{zhuang2020comprehensive, saito2018maximum} is a promising technique for deep learning models to transfer general knowledge across similar machine learning tasks. Domain-adversarial neural network (DANN) \cite{ganin2016domain} is proposed to learn a feature extractor to extract domain-independent features in an adversarial regime. This approach and its variants have been applied to many research fields, such as image classification \cite{ganin2016domain,tzeng2017adversarial} and semantic segmentation \cite{tsai2018learning}.

In this paper, we propose a deep domain adversarial adaptation method with an effective network structure to tackle the domain shift problem in photon-efficient imaging. This method only requires dozens of extra training iterations with a small amount of unlabeled data from the target domain for the model to adapt to different SBR or hardware-specific properties. Extensive experiments on the simulated datasets demonstrate the efficiency of our method for the model to adapt to different SBRs. Moreover, the real-world experiment based on our home-built single-photon imaging system shows that our method improves the baseline model performance by a remarkable margin, while other statistical and deep learning algorithms do not generalize well to the realistic scenario. 

\section{Method}
\begin{figure}[]
	\centering
	\includegraphics[width=0.48\textwidth]{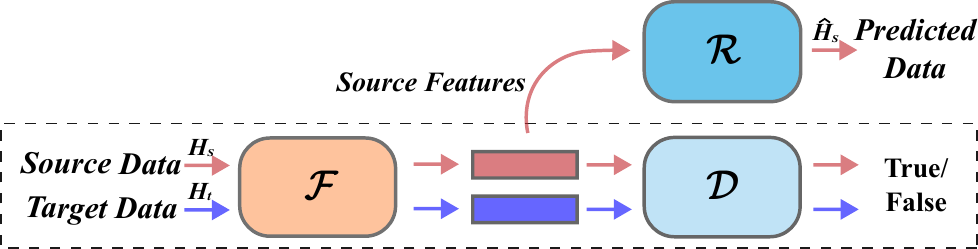}
	\caption{Schematic diagram of the deep domain adversarial adaptation scheme. $\mathcal{F}$, feature extractor; $\mathcal{D}$, discriminator; $\mathcal{R}$, reconstructor. We use red and blue colors to indicate the pipelines on the source and target domain, respectively.}
	\label{Fig:pipeline}
\end{figure}

The domain adversarial adaptation method is adopted mainly from \cite{ganin2016domain} and its schematic diagram is shown in \fref{Fig:pipeline}. The feature extractor, denoted by $\mathcal{F}$, learns the transformation that maps both domains into a common feature space. The purpose of this transformation is to extract domain-independent features from the inputs to confuse the discriminator, while the discriminator, denoted by $\mathcal{D}$, tries to determine whether these features are from the source or the target domain. This process, as marked by the dashed box, corresponds to a minimax two-player game which forces $\mathcal{F}$ to focus on capturing the similarity of the two domains. Meanwhile, the reconstructor denoted by $\mathcal{R}$ learns to estimate the ground-truth depth from the source features.

In principle, this scheme can be applied to any deep learning model as long as a suitable feature space has been defined. Nevertheless, existing deep learning models either involve complex attention operations \cite{peng2020photon} or residual connections \cite{zang2021non}, which makes it difficult to find an encoder-decoder division for adversarial adaptation training. Therefore, we propose a simple yet efficient encoder-decoder network structure design for the domain adaptation task. Firstly, inspired by \cite{szegedy2016rethinking} and \cite{christoph2016spatiotemporal}, we design a spatiotemporal block (ST-Block) which consists of four branches based on the spatiotemporal convolution (see \Aref{App:STIN-1} for more details). $\mathcal{F}$ is built by stacking eight modules, each composed of an ST-Block and a 3D pooling layer. Then $\mathcal{R}$ comprises seven deconvolutional layers and a $1 \times 1 \times 1$ convolutional layer. In particular, each convolutional and deconvolutional layer is followed by a group normalization layer and a rectified linear unit (ReLU) activation. We name the whole composition of $\mathcal{F}$ and $\mathcal{R}$ as the spatiotemporal inception network (STIN). The structure of STIN is given in \Aref{App:STIN-2}.

\subsection{Pre-training}
STIN is first trained on the source domain as a baseline model. Here the depth estimation is taken as a classification problem, in which STIN aims at correctly classifying each pixel into a depth category that matches its ground-truth depth \cite{yao2022robust}. The photon counting histograms $H \in \mathbb{R}^{T \times N_x \times N_y}$ for the source domain are generated by simulation from the ground-truth depth images ${Z} \in \mathbb{R}^{N_x \times N_y}$, where $T$, $N_x$ and $N_y$ are the number of time bins, the horizontal and vertical resolutions, respectively. The prediction of the model is denoted as $\hat{H} =\mathcal{R} (\mathcal{F}(H)) \in \mathbb{R}^{T \times N_x \times N_y}$. Note that $Z$ has to be further processed by the following rounding function for classification
\begin{equation}
	z_{i,j} = [z_{i,j}]=\max \{n \in \mathbb{Z}^{+}_{0} \mid n \leq \dfrac{ 2z_{i,j}}{\Delta \cdot {\rm c}} \},
\end{equation}
where ${\rm c}$ is the light speed and $\Delta$ is the size of time bin. The cross-entropy (CE) loss between $\hat{H}$ and $Z$ is given by
\begin{equation}
	\mathcal{L}_{\mathrm{CE}}(\hat{{H}}, {Z}) =\sum_{i,j} \mathcal{L}_{\mathrm{CE}} (\bm{\hat{h}_{i,j}}, \bm{z^{({\rm o})}_{i,j}} )
	\label{Eq:loss_ce}
\end{equation}
which is adopted from \cite{yao2022robust}, and $\bm{z^{({\rm o})}_{i,j}} \in \mathbb{R}^{T}$ is the one-hot encoding of $z_{i,j}$. Another total variation (TV)  regularization term $\mathcal{L}_{ \mathrm{TV}}$ for smoothing the prediction \cite{zang2021non} is added to make the total loss function as 
\begin{align}
	\mathcal{L} = \mathcal{L}_{ \mathrm{CE}} + \lambda_{\mathrm{TV}} \cdot \mathcal{L}_{ \mathrm{TV}},
	\label{Eq:losstotal}
\end{align}
where $\lambda_{\mathrm{TV}}$ is the weight of the TV term. The details of $\mathcal{L}_{\mathrm{CE}}$ and $\mathcal{L}_{\mathrm{TV}}$ are given in \Aref{App:STIN-3}. $\mathcal{F}$ and $\mathcal{R}$ are trained by standard backpropagation with respect to the loss \eref{Eq:losstotal} on the source domain.

\subsection{Domain adversarial training}

\begin{figure}[t]
	\renewcommand{\algorithmicrequire}{\textbf{Input:}}
	\renewcommand{\algorithmicensure}{\textbf{Output:}}
	\begin{algorithm}[H]
		\caption{Implementation of DANN}
		\label{Alt:imple}
		\begin{algorithmic}[1]
			\Require \qquad\\
			samples: $\bm{\mathrm{S}}=\{({H_s}^{(i)}, {Z_s}^{(i)})\}_{i=1}^{N}$ and \\
			\ \ \ \qquad \quad  $\bm{\mathrm{T}}=\{{H_t}^{(i)}\}_{j=1}^{N^{\prime}}$; \\
			adaptation weight $\lambda_{\mathrm{a}}$;\\
			pre-trained $\mathcal{F}$ and $\mathcal{R}$ on the source domain;\\
			randomly initialized $\mathcal{D}$;
			\Ensure new $\mathcal{F}$ and $\mathcal{R}$;
			\While {stopping criterion is not met}
			\For{$i \leftarrow 1, \cdots, N$}
			\State Randomly sample $j$ from $\{1,\cdots,N^{\prime}\}$;
			\State Calculate $\mathcal{L}_{\operatorname{adv}}({H^{(i)}_{s}}, {H^{(j)}_{t}})$;
			\State Update $\mathcal{D}$ by minimizing $\mathcal{L}_{\operatorname{adv}}$;
			\State  Calculate the total loss by \\$ \begin{aligned}
				\qquad \qquad \qquad	& \mathcal{L}_{\operatorname{total}}({H^{(i)}_{s}}, {Z^{(i)}_{s}}, {H^{(j)}_{t}}) \leftarrow \\
				&\mathcal{L}({H^{(i)}_{s}}, {Z^{(i)}_{s}}) - \lambda_{\mathrm{a}} \cdot \mathcal{L}_{\operatorname{adv}}({H^{(i)}_{s}}, {H^{(j)}_{t}});
			\end{aligned}$
			\State Update $\mathcal{F}$ and $\mathcal{R}$ by minimizing $\mathcal{L}_{\operatorname{total}}$;
			\EndFor
			\EndWhile
		\end{algorithmic}	
	\end{algorithm}
	\vspace{-5pt}
\end{figure}

Domain adversarial training is achieved by alternatively minimizing the following loss functions,
\begin{align}
	&\mathcal{L}(H_{s}, Z_{s})=\mathcal{L} (\mathcal{R}(\mathcal{F}(H_{s})), Z_{s} ), \label{eq:loss_src}\\
	& \mathcal{L}_{\operatorname{adv}}(H_{s}, H_{t})=\log \mathcal{D}(\mathcal{F}({H_{s}})) - \log (1-\mathcal{D}(\mathcal{F}({H_{t}})), \label{eq:loss_adv}
\end{align}
where $H_{s}$ and $Z_{s}$ are the sample and ground-truth depth from the source domain, and $H_{t}$ is the sample from the target domain. It should be noted that the ground-truth depth of the target domain is not needed for domain adversarial training. In each training iteration, $\mathcal{D}$ is firstly optimized by minimizing the domain adversarial loss defined in \eref{eq:loss_adv} to extract domain-invariant features. Then $\mathcal{F}$ and $\mathcal{R}$ are simultaneously optimized by minimizing the total loss function given by
\begin{equation}
	\mathcal{L}_{\operatorname{total}} =  \mathcal{L} - \lambda_{\mathrm{a}} \cdot \mathcal{L}_{\operatorname{adv}}
\end{equation}
to improve the reconstruction performance, with $\lambda_{\mathrm{a}}$ being the weight of adaptation. The detailed implementation of DANN training is shown in \Alref{Alt:imple}, in which we denote $\mathrm{S}$ and $\mathrm{T}$ as the training sets on the source and target domain, respectively.

The goal of the discriminator is to distinguish between the source and target domain from the latent features. Thus, ${Z}$ has to maintain the temporal dimension of the features, while the spatial dimension of the features needs to be blurred. To do this, the latent features first go through a 3D pooling layer with a kernel size of $ 1\times 8\times 8$. After that, three fully-connected layers will map these features to a scalar that classifies the input into either the source or the target domain.

 After the adversarial training, the target depth image denoted by $\hat{Z}_{t}$ is predicted by the Softargmax function \cite{lindell2018single} as
\begin{align}
	[\hat{z}_{t}]_{i,j} &= \frac{\Delta \cdot {\rm c}}{2} \operatorname{Softargmax}(\bm{[\hat{h}_{t}]_{i,j}}) \nonumber\\
	&=\frac{\Delta \cdot {\rm c}}{2} \sum_{k=1}^{T} k \cdot [\hat{h}_{t}]_{k,i,j},
	\label{Eq:softargmax}
	\vspace{-5pt}
\end{align}
where $\hat{H}_t = \mathcal{R}(\mathcal{F}(H_t))$. Note that the Softargmax function is equivalent to maximum likelihood estimation (MLE) \cite{myung2003tutorial}.

\section{Results}
\subsection{Metrics and implementation details}
The following metrics are used for the quantitative evaluation:
\begin{itemize}
	\item Root mean squared error (RMSE), defined as 
	\begin{equation}
		{\rm RMSE}(Z, \hat{Z})=\sqrt{\frac{1}{N_x N_y} \sum_{i}^{N_x} \sum_{j}^{N_y}\left(z_{i,j}-\hat{z}_{i,j}\right)^{2}}
	\end{equation}
	\item Absolute relative difference (Abs rel), defined as
	\begin{equation}
		{\rm Abs \ rel} (Z, \hat{Z})=\frac{1}{N_x N_y} \sum_{i}^{N_x} \sum_{j}^{N_y} \frac{\left|z_{i,j}-\hat{z}_{i,j}\right|}{z_{i,j}}
	\end{equation}
\end{itemize}

All the models are implemented using the deep learning library Pytorch \cite{paszke2019pytorch} and trained on a single GTX 1080ti GPU. We adopt the Adam optimizer \cite{AdamKingma2015} with a base learning rate of $0.001$. $\lambda_{\mathrm{TV}}$ and $\lambda_{\mathrm{a}}$ are set to be $0.001$ and  $0.1$, respectively. The code for implementing our model is available at \cite{codeurl}.

\subsection{Simulated experiments}
The validation experiments are firstly conducted on the simulated dataset. The goal is to adapt the model pre-trained on low-noise domain to high-noise domain. $15940$ low-noise samples with ground-truth depth information and $42$ high-noise samples without ground-truth information are generated from NYU~v2 dataset \cite{silberman2012indoor} as the training set using the observation model described in \Aref{App:Observation}. The total number of time bin is $1024$ and the bin size is $80$ picoseconds (ps). The waveform of the emitted pulsed is assumed to have an ideal Gaussian shape with the full width at half maximum of $400$ ps. Constrained by GPU memory size, we divide the training samples into patches of size $[1024, 32, 32]$ and set the batch size to $6$. The SBRs for generating the low-noise and high-noise samples are \{2:2, 5:2, 10:2\} and \{2:50, 2:100\}, respectively. Here an SBR of $n$:$m$ stands for the ratio of the average number of signal photons to noise photons per pixel. The test set is composed of samples with SBRs of \{2:2, 2:50, 2:100\}, which are generated from Middlebury dataset \cite{scharstein2007learning} with a resolution of $512 \times 512$. As noted before, the STIN is first trained on the low-noise dataset as the baseline model. The training takes about 10 hours. Then the trained model is adapted to the high-noise domain using the $42$ high-noise samples by domain adversarial training, which takes only about 10 minutes. In testing, the high resolution input image is divided into small patches for depth prediction and then the outputs are concatenated into a large image. 

As shown in \tref{tab:result-da}, the baseline model does not generalize well to the high-noise test samples with SBR of 2:50 or 2:100, whereas the DANN-based adaptation method achieves a significant improvement at low SBRs. Particularly, the DANN-based model outperforms the baseline by $77.3\%$ in RMSE and $72.5\%$ in Abs rel when SBR is 2:100. At the high SBR of 2:2, the performance of DANN-based model is worse than the baseline model, since the baseline is trained solely with this SBR. Meanwhile, the DANN-based model has to adapt to high-noise samples, which inevitably decreases its performance at SBR=2:2 since the model needs to balance between two domains.

\begin{table}[]
	\renewcommand\arraystretch{1.2}
	\centering
	\caption{Comparisons of the baseline and DANN-based model on simulated datasets. Best results are highlighted in bold. The percentage difference from the baseline is also shown in the parentheses. }
	
	\begin{tabular}{c|cc|cc}
		\hline
		& \multicolumn{2}{c|}{Baseline} & \multicolumn{2}{c}{DANN-based STIN}  \\ \hline
		SBR& RMSE       & Abs rel     & RMSE        & Abs rel     \\ \hline
		2:2     & $\bm{0.026}$& $\bm{0.018}$&$0.031 (\uparrow 19.6 \%)$&$ 0.019(\uparrow 3.8 \%)$ \\ \hline
		2:50    & $0.121$& $0.036$&$ \bm{0.076(\downarrow 37.1\%)}$& $\bm{0.028(\downarrow 24.9\%)}$  \\ \hline
		2:100   & $0.388$& $0.147 $& $\bm{0.088(\downarrow 77.3\%)} $& \bm{$0.040(\downarrow 72.5\%)}$ \\ \hline
	\end{tabular}
	\label{tab:result-da}
\end{table}

\begin{figure}[]
	\centering
	\includegraphics[width=0.45\textwidth]{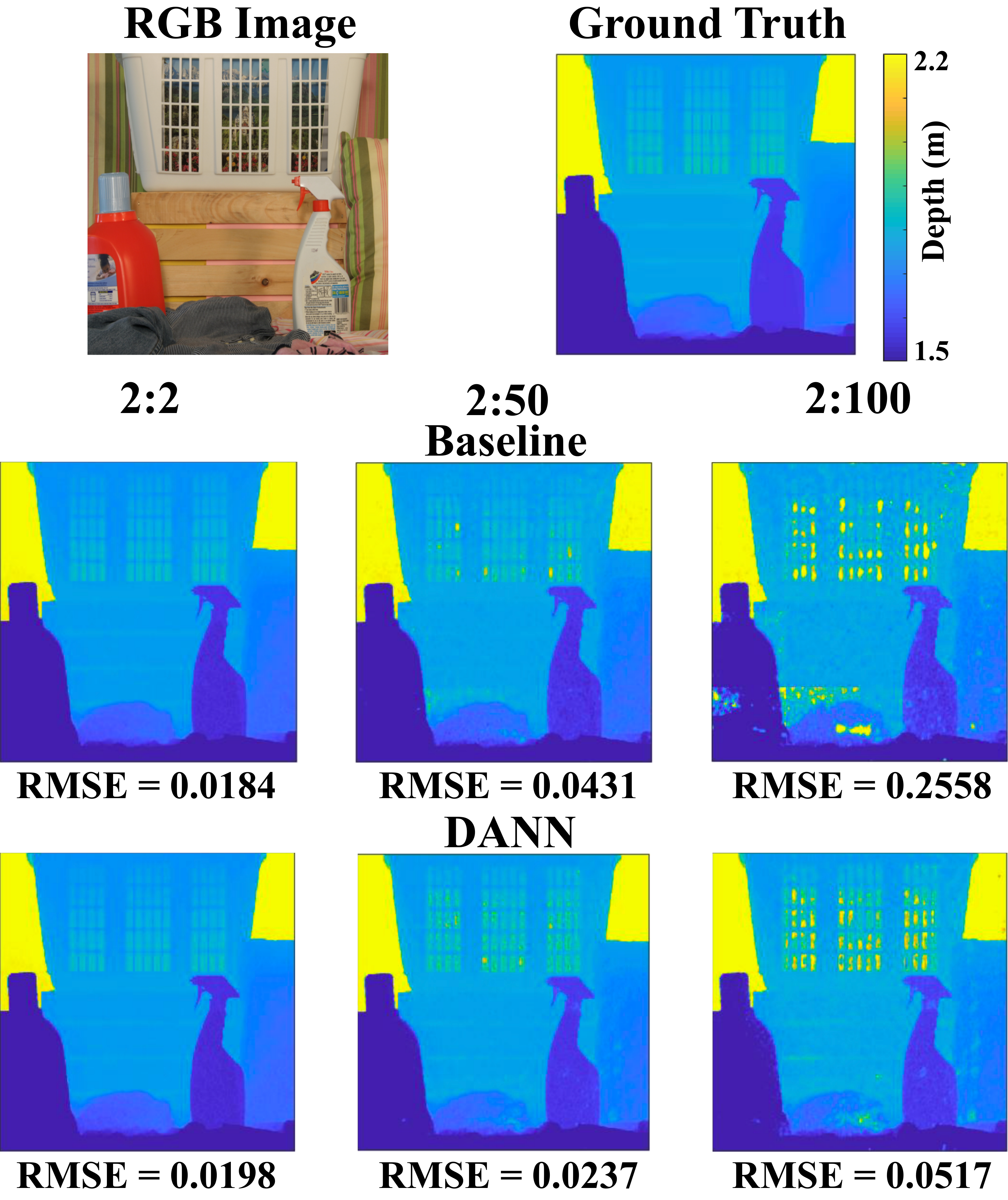}
	\caption{The results of a random selected scene named Laundry with different SBRs in the Middlebury dataset. The first row includes the RGB image and ground truth. The second and third row include the depth images predicted by STIN and DANN-based STIN, respectively.}
	\label{fig:result-da}
\end{figure}

\fref{fig:result-da} shows the predicted depth images for the Laundry scene from the test set. We see that the baseline can precisely recover the depth image at high SBRs, whereas many erroneous regions appear when SBR decreases. In contrast, the performance of the DANN-based model is much more robust to the decrease in SBR.

\begin{table}[]
	\renewcommand\arraystretch{1.2}
	\centering
	\caption{RMSEs for different $\lambda_{\mathrm{TV}}$ and $\lambda_{\mathrm{a}}$ on the simulated dataset with the SBR of 2:50. The best result is highlighted in bold. }
	\label{tab:lambda}
	\begin{tabular}{c|c|c|c}
		\hline
		&  $\lambda_{\mathrm{TV}} = 0.01$ & $\lambda_{\mathrm{TV}} = 0.001$ & $\lambda_{\mathrm{TV}} = 0.0001$\\ \hline
		$\lambda_{\mathrm{a}} = 1$     & $0.078$& $0.077$&$0.085$\\ \hline
		$\lambda_{\mathrm{a}} = 0.1$    & $0.078$& $\bm{0.076}$&$ 0.082$\\ \hline
		$\lambda_{\mathrm{a}} = 0.01$   & $0.113$& $0.119 $& $0.114 $\\ \hline
	\end{tabular}
\end{table}

We have conducted extra experiments to study the sensitivity of $\lambda_{\mathrm{TV}}$ and $\lambda_{\mathrm{a}}$; see \tref{tab:lambda}. When the weight of adaptation is too small, such as $\lambda_{\mathrm{a}} = 0.01$, the improvement in performance is not significant. Otherwise, the protocol is not sensitive to the regularization coefficient $\lambda_{\mathrm{TV}}$.

\subsection{Real-world experiments}

\begin{figure}[]
	\centering
	\includegraphics[width=0.48\textwidth]{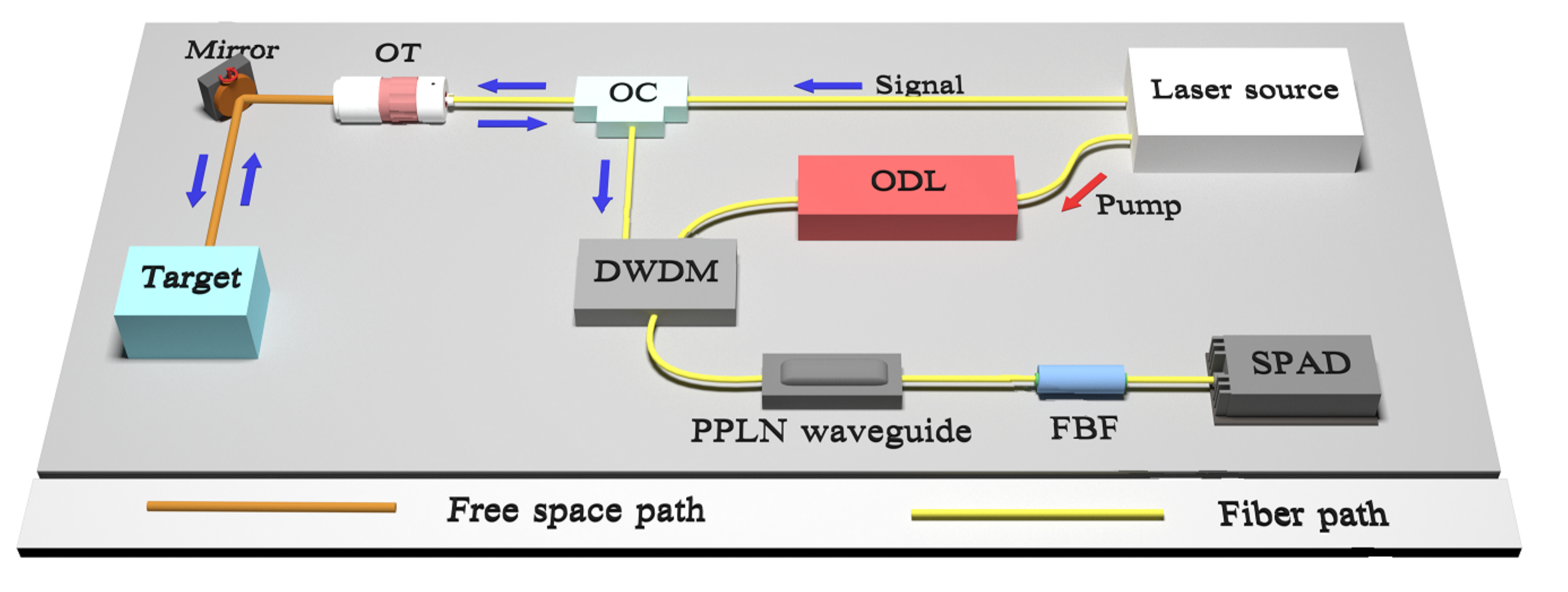}
	\caption{Schematic diagram of our imaging system. ODL, optical delay line; OC, optical circulator; OT, optical transceiver; DWDM, dense wavelength division multiplexing; PPLN waveguide, periodically poled lithium niobate waveguide; FBF, fiber bandpass filter; SPAD, single-photon avalanche diode.}
	\label{Fig:system}
\end{figure}

Next, we conduct real-world experiment by building a single-photon imaging system, whose schematic diagram is shown in \fref{Fig:system}. The system is based on an upconversion process that transforms the near-infrared signal photon to the visible regime for detection \cite{rehain2020noise}. The picosecond laser source emits synchronized pump pulses (1565.5 nm) and signal pulses (1554.1 nm) with the repetition rate of 600 MHz. The signal pulses are propagated to the programmable gimbal-less two-axis micro-electro-mechanical system (MEMS) mirror through an optical transceiver for scanning. In the meantime, the pump pulses are delayed in the programmable motorized ODL.  By adjusting the delay time such that the signal and pump pulses coincide (the optical delay time matches the target distance), the upconversion light of 779.8 nm for photon detection is generated by the sum frequency generation process in the PPLN waveguide. Then the generated light is passed through several home-built FBFs to filter out the noise, before finally being detected by the SPAD detector. We use a field-programmable gate array (FPGA) to count the number of detected photons at each time-delay step. A desktop computer is employed to control the MEMS mirror and ODL. Unlike \cite{rehain2020noise}, we use commercially packaged PPLN waveguide instead of bulk PPLN crystal in the conversion module. Compared with bulk crystal, the waveguide has higher upconversion efficiency with a more stable all-fiber structure.

\begin{figure}[]
	\centering
	\includegraphics[width=0.45\textwidth]{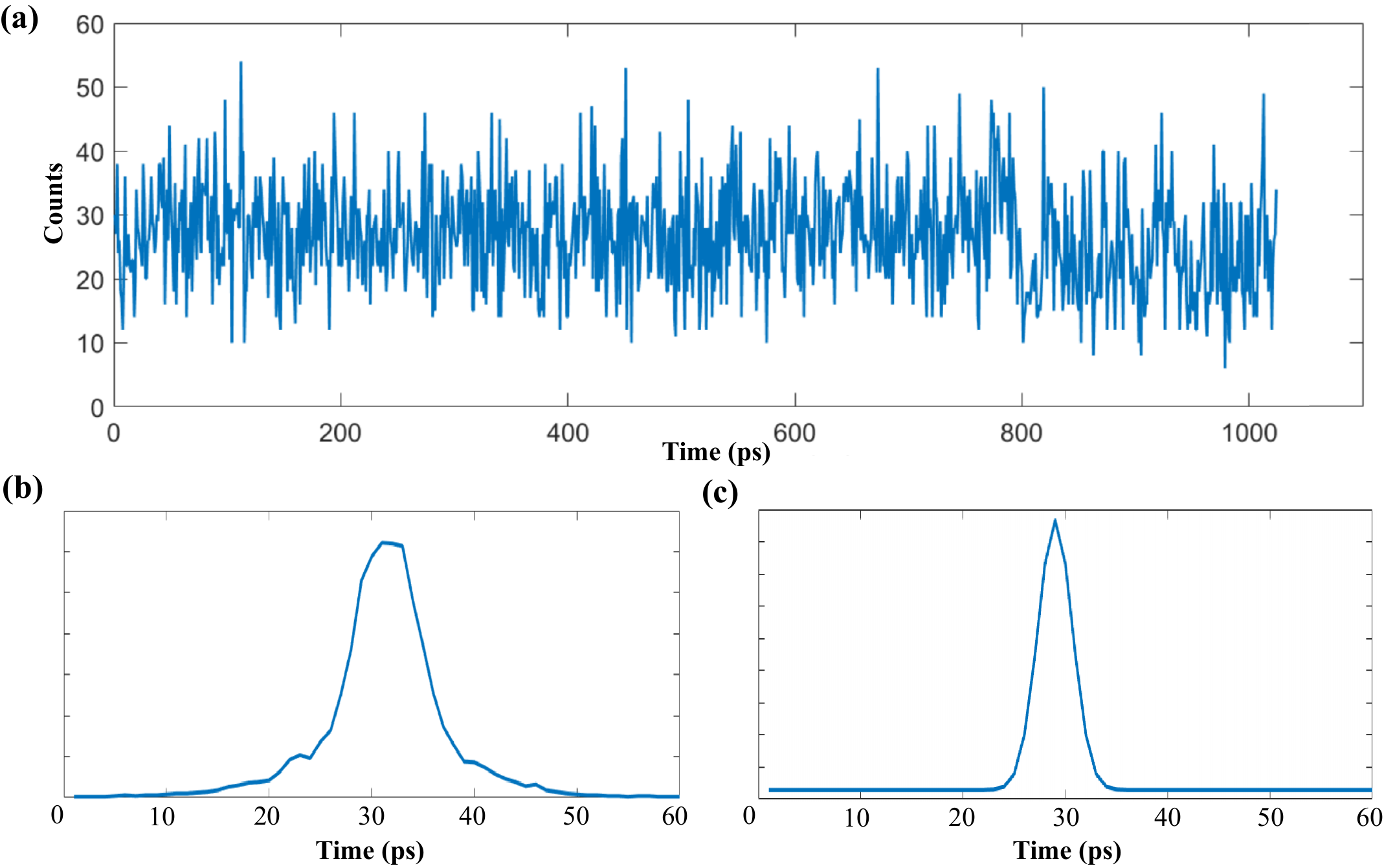}
	\caption{(a) Noise photon counts of our system. (b) The normalized real-world pulse shape of our system. (c) The normalized ideal pulse shape used for simulation.}
	\label{Fig:waveform}
\end{figure}

The detected noise photons as shown in \fref{Fig:waveform} (a) come from the environment, dark counts of the SPAD and the pump light. In particular, the frequency-doubled photons of the pump light that cannot be completely eliminated by the optical filters constitute the major source of noise, whereas the simulation only assumes a uniform distribution for ambient noise. Besides, it can be observed that there is a visible difference between the real-world waveform of detected photons in \fref{Fig:waveform} (b) and the simulated waveform in \fref{Fig:waveform} (c). Therefore, the noise source, statistics and detected waveform of our system substantially differ from the ideal simulation, which makes it a perfect candidate for the validation of the domain adaptation method.

\begin{figure}[]
	\centering
	\includegraphics[width=0.45\textwidth]{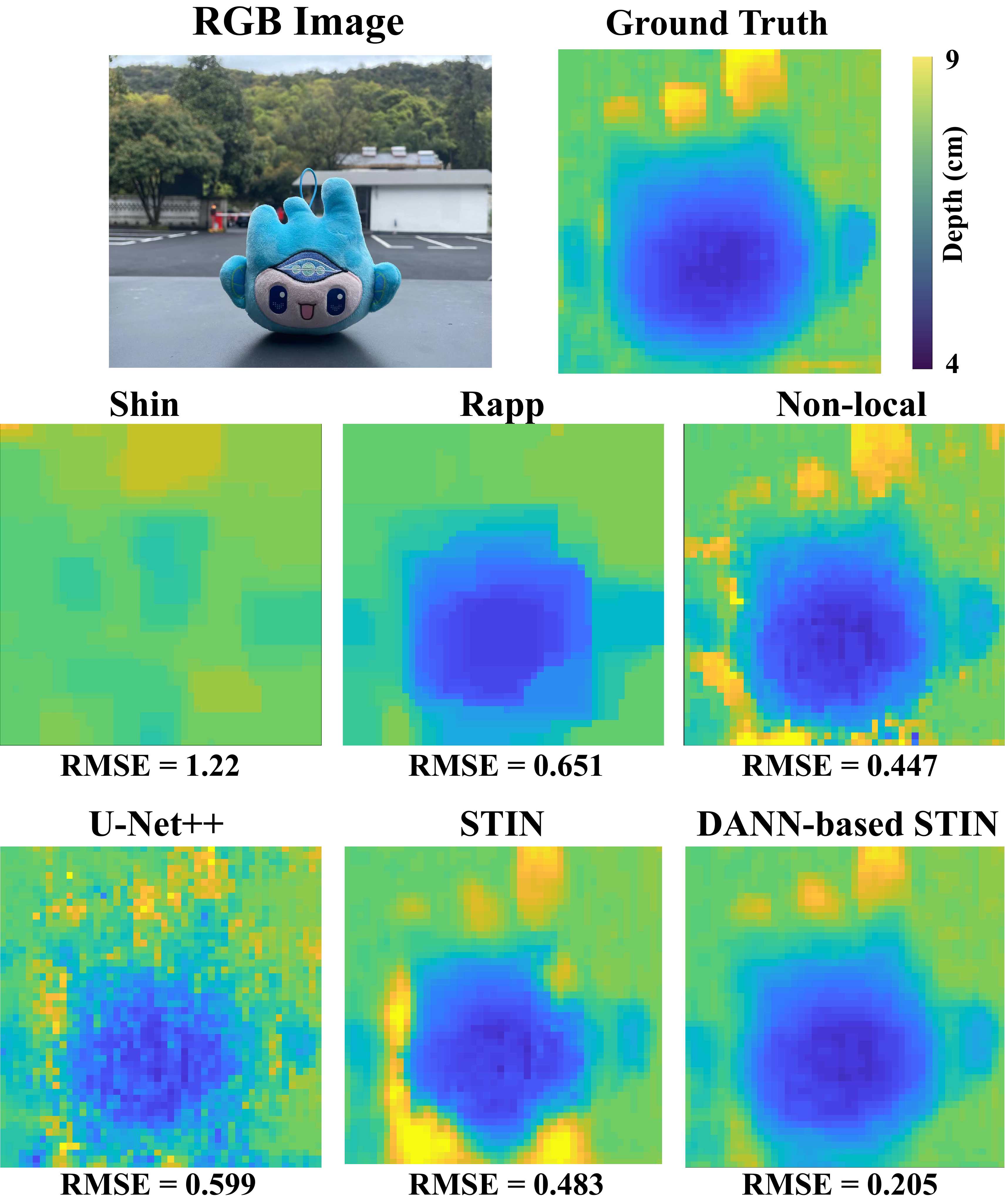}
	\caption{The reconstruction results of 2022 Asian Games mascot Chenchen with a spatial resolution of 48 $\times$ 48. The estimated ground-truth data are only used for evaluation but not for adaptation.}
	\label{fig:result}
\end{figure}

\begin{table}[]
	\renewcommand\arraystretch{1.2}
	\centering
	\caption{The comparison with respect to RMSE, Abs rel and Accuracy with $\delta$ (defined in \Aref{App:Results}) of the deep learning methods on the real-world experiment. Best results are highlighted in bold.}
	\begin{tabular}{c|c|c|c|c}
		\hline
		&\multicolumn{2}{c|}{Error} & \multicolumn{2}{c}{Accuracy with $\delta$ }\\ \hline
		Models &    RMSE (cm)   &    Abs rel    &$\delta = 1.01$   & \ $\delta = (1.01)^2$  \\ \hline
		Non-local &      $0.447$      &  $0.0015$    &  $33.3\%$     &   $53.9\%$   \\ \hline
		U-Net++ &      $0.599$      &  $0.0021$    &  $19.5\%$     &   $30.3\%$   \\ \hline
		STIN &      $0.483$      &  $0.0012$    &  $30.5\%$     &   $61.3\%$   \\ \hline
		DANN &      $\bm{0.205}$      &  $\bm{0.0008}$     &  $\bm{40.8\%}$      &   $\bm{66.3\%}$    \\ \hline
	\end{tabular}
	\label{tab:quan}
\end{table}
The 2022 Asian Games mascot Chenchen is placed outdoors and about 5 meters away from the optical transceiver. The ODL is scanned in 1 ps steps over a range of 1024 ps. In order to simulate photon-efficient imaging, we set the acquisition time for each time delay step as 1 microsecond such that no more than dozens of signal and noise photons are recorded at each pixel. The baseline STIN is pre-trained on the simulated dataset with SBRs of \{2:10, 2:50, 2:100\}. By adjusting the number of optical filters, the overall SBR of our system is measured to be 0.103, which does not overlap with the SBRs for pre-training. The pre-trained STIN is then tuned by several patches with the resolution of $[1024,32,32]$ from the real-world data using the proposed domain adversarial adaptation method. The domain adaptation process takes only several minutes, without using any ground-truth depth information in the real world. In evaluation, by increasing the acquisition time at each time-delay step to one millisecond like \cite{peng2020photon}, we collected enough signal photon counts for the precise depth estimation of each pixel that approximates the ground truth using MLE. As shown in \fref{fig:result}, our method achieves a significant visual improvement when compared to other state-of-the-art competitors, including two statistical methods \cite{shin2016photon, rapp2017few}, two deep learning models \cite{peng2020photon, zang2021non} and the baseline model. These competitors all have shown good performance on the simulated dataset as detailed in \Aref{App:Results} and poor generalizability in real experiment, which highlights the necessity of domain adaptation. We have provided a more comprehensive comparison in \tref{tab:quan}. In particular, the DANN-based STIN achieves the highest imaging quality with a decrease in RMSE of $57.5\%$, $65.8\%$ and $54.1\%$ and in Abs rel of $33.3\%$, $61.9\%$ and $46.6\%$, as compared to baseline, U-Net++ and Non-local net, respectively. In addition, DANN achieves significant improvements in the metric Accuracy with $\delta$.

\section{Conclusion}
We propose an efficient domain adversarial adaptation method to improve the depth estimation precision for a 3D scene, without using any ground-truth depth information of the target domain. Experimental results on simulated dataset and our home-built imaging system have verified the effectiveness of our method. It has been shown that although the pre-trained models may be effective on the simulated dataset, their performance may deteriorate significantly if domain adaptation is not conducted. Considering that obtaining sufficient data, either real-world data with ground-truth labels or simulated data, to retrain the model in the target domain is extremely time-consuming or merely impossible, the proposed approach of this paper is able to provide significant resource savings for the practical deployment of deep learning models in photon-efficient imaging applications. 

\begin{acknowledgments}
This work was supported by the National Natural Science Foundation of China (Grant No. 62173296).
\end{acknowledgments}

\appendix

\section{Details of STIN}
\subsection{ST-block} 
\label{App:STIN-1}
The first branch consists of a $1 \times 1 \times 1$ 3D convolutional layer. The second branch is composed of a $1 \times 1 \times 1$ convolutional layer followed by an $n_t \times 1 \times 1$ and a $1 \times n_s \times n_s$ 3D convolutional layer. Here $n_t$ and $n_s$ are the kernel size for the temporal and spatial dimensions, respectively. The third branch swaps the locations of the temporal and spatial convolutions in the second branch. The fourth branch is composed of a $1 \times 1 \times 1$ convolutional layer, two $1 \times n_s \times n_s$ convolutional layers and an $n_t \times 1 \times 1$ convolutional layer. Finally, the outputs of four branches are concatenated together along the channel dimension. For example, if the size of the input is $[B,C_i,T,H,W]$, then the size of the final output is $[B, 4 \times C_{b}, T, H, W]$, where $C_{b}$ is the number of output channels on each branch. 

\subsection{Network structure} 
\label{App:STIN-2}
We summarize the detailed structure of the feature extractor, the discriminator and the reconstructor in \tref{tab:fe}, \tref{tab:d} and \tref{tab:r}, respectively.

\begin{table}[]
	\renewcommand\arraystretch{1.1}
	\centering
	\caption{The structure of the feature extractor. }
	\begin{tabular}{c|c|c}
		\hline
		Layer    & Kernel size &  Output size  \\ \hline
		Input    & - & $[1,1024,32,32]$ \\ \hline
		ST-block &  $7 \times 3 \times 3$ &  $[4,1024,32,32]$ \\ \hline
		Max pooling    &  $2 \times 1 \times 1$ &  $[4,512,32,32]$    \\ \hline
		ST-block &   $7 \times 3 \times 3$ &  $[8,512,32,32]$  \\ \hline
		Max pooling    &   $2 \times 1 \times 1$&  $[8,256,32,32]$    \\ \hline
		ST-block &     $7 \times 3 \times 3$ &  $[12 ,256 ,32 , 32]$      \\ \hline
		Max pooling    &       $2 \times 1 \times 1$&  $[12 , 128 , 32 , 32]$   \\ \hline
		ST-block &      $7 \times 3 \times 3$ &  $[16 , 128 , 32 , 32]$      \\\hline
		Max pooling    &     $2 \times 1 \times 1$&  $[16 , 64 , 32 , 32]$    \\\hline
		ST-block &        $7 \times 3 \times 3$&  $[24 , 64 , 32 , 32]$    \\\hline
		Max pooling    &     $2 \times 1 \times 1$&  $[24 , 32 , 32 , 32]$   \\\hline
		ST-block &     $7 \times 3 \times 3$ &  $[32 , 32 , 32 , 32]$     \\\hline
		Max pooling    &     $2 \times 1 \times 1$&  $[32 , 16 , 32 , 32]$   \\\hline  
		ST-block &        $7 \times 3 \times 3$ &  $[40 , 16 , 32 , 32]$    \\\hline
		Max pooling    &     $2 \times 1 \times 1$&  $[40 , 8 , 32 , 32]$   \\\hline  
		ST-block &        $7 \times 3 \times 3$&  $[48 , 8 , 32 , 32]$    \\\hline
	\end{tabular}
	\label{tab:fe}
\end{table}

\begin{table}[htbp]
	\renewcommand\arraystretch{1.1}
	\centering
	\caption{The structure of the discriminator. FC denotes the fully-connected layer.}
	\begin{tabular}{c|c|c}
		\hline
		Layer    & Kernel size & Output size  \\ \hline
		Input    & - &  $[48 , 8 , 32 , 32]$ \\ \hline
		Average pooling &  $1\times 8 \times 8$&  $[48 , 8 , 4 , 4]$ \\ \hline
		Reshape &  -&  $[6144 ]$ \\ \hline
		FC &  - &  $[512]$ \\ \hline
		FC &  - &  $[128]$ \\ \hline
		FC &  - &  $[1]$ \\ \hline
	\end{tabular}
	\label{tab:d}
\end{table}

\begin{table}[]
	\renewcommand\arraystretch{1.1}
	\centering
	\caption{The structure of the reconstructor.}
	\begin{tabular}{c|c|c}
		\hline
		Layer    & Kernel size  & Output size  \\ \hline
		Input    & - & $[48, 8,32, 32]$ \\ \hline
		3D Deconv &  $6\times 3 \times 3$   &  $[40 , 16 , 32 , 32]$ \\ \hline
		3D Deconv &  $6\times 3 \times 3$ &  $[32 , 32 , 32 , 32]$ \\ \hline
		3D Deconv &  $6\times 3 \times 3$ &  $[24 , 64 , 32 , 32]$ \\ \hline
		3D Deconv &  $6\times 3 \times 3$ &  $[16 , 128 , 32 , 32]$ \\ \hline
		3D Deconv &  $6\times 3 \times 3$ &  $[12 , 256 , 32 , 32]$ \\ \hline
		3D Deconv &  $6\times 3 \times 3$ &  $[8 , 512 , 32 , 32]$ \\ \hline
		3D Deconv &  $6\times 3 \times 3$&  $[4 , 1024 , 32 , 32]$ \\ \hline
		3D Conv &  $1\times 1 \times 1$&  $[1 , 1024 , 32 , 32]$ \\ \hline
	\end{tabular}
	\label{tab:r}
\end{table}

\subsection{The loss functions}
\label{App:STIN-3}
The CE loss for $\bm{a}\in \mathbb{R}^{T}$ and $\bm{b}\in \mathbb{R}^{T}$ is defined as
\begin{equation}
	\mathcal{L}_{\mathrm{CE}}(\bm{a}, \bm{b})=-\sum_{t}^{T} a_{t} \log b_{t}.
\end{equation}

The TV loss for the input $Z \in \mathbb{R}^{N_x\times N_y}$ is defined as
\begin{equation}
	\mathcal{L}_{ \mathrm{TV}}({\hat{Z}})= \sum_{i, j}(|\hat{z}_{i+1, j}-\hat{z}_{i, j}|+|\hat{z}_{i, j+1}-\hat{z}_{i, j}|).
	\vspace{-10pt}
\end{equation}

\begin{table*}[]
	\renewcommand\arraystretch{1.1}
	\centering
	\caption{Comparison results on the simulated dataset. Best results are highlighted in bold.}
	\label{Tab:results}
	\begin{tabular}{c|cc|ccc}
		\hline
		\multicolumn{6}{c}{SBR = 2:10}                            \\ \hline
		& \multicolumn{2}{c}{Error (Lower is better)} & \multicolumn{3}{c}{Accuracy with $\delta$ (Higher is better)} \\ \hline
		Models &    RMSE(m)   &    Abs rel    &$\delta = 1.01$   & \ $\delta = (1.01)^2$  &$\delta =  (1.01)^3$   \\ \hline
		Shin &      $3.2225$      &  $3.1995$    &  $0$     &   $0$    & $0$  \\
		Rapp &        $ 0.0610$      &  $0.0350$    &    $43.45\%$   &     $71.05\%$  &$82.74\%$   \\ \hline
		Non-local Net  &  $0.0244$ &  $0.0106$        &     $55.30\%$     &     $93.68\%$  &  $97.58\%$    \\
		U-Net++ &   $ 0.0278$  &  $\bm{0.0087}$   &      \bm{$78.79\%$}     &     $93.46\%$  &  $96.22\%$\\
		STIN&   $\bm{0.0211}$    &   $ 0.0094$      &       $62.15\%$     &     \bm{ $94.76\%$ }&  \bm{ $99.35\%$ }   \\\hline
		\multicolumn{6}{c}{SBR = 2:50}                           \\\hline
		Shin &    $4.1783$     &    $4.1723$    &   $0$    &  $0$     &   $0$  \\
		Rapp &     $0.0676$   &   $0.0399$       &   $39.87\%$    &   $67.71\%$    & $79.42\%$  \\ 
		Non-local Net &   $0.0297$ &   $ 0.0113$       &       $55.58\%$     &     $91.66\%$  &  $96.16\%$   \\
		U-Net++ &   $0.0787$ &  $ 0.0174$       &     \bm{$70.96\%$}  &    $85.11\%$    &   $88.84\%$  \\ 
		STIN&  $\bm{0.0249}$ &    $\bm{0.0102}$     &       $60.01\%$     &     \bm{ $92.43\%$ } & \bm{  $97.08\%$  }   \\\hline
		\multicolumn{6}{c}{SBR = 2:100}                            \\\hline
		Shin &     $4.2955$    &      $4.2901$     &    $0$   &   $0$    & $0$\\
		Rapp  &   $0.0726$     &       $ 0.0442$    &       $38.08\%$   &   $64.83\%$    & $76.61\%$   \\ \hline
		Non-local Net  & $0.0329$ & $ 0.0122$         &       $52.66\%$     &     $89.12\%$  &  $95.05\%$   \\
		U-Net++ &  $0.0795$ &  $ 0.0172$       &  \bm{$70.57\%$}     &     $85.36\%$  &  $89.16\%$ \\ \hline
		STIN&   $\bm{0.0305}$ &   $\bm{0.0112}$       &   $57.92\%$    &    \bm{ $90.38\%$ }  &   \bm{ $95.66\%$ }  \\\hline
	\end{tabular}
\end{table*} 

\section{Further Results on simulated dataset}
\label{App:Results}
The comparison of Shin \cite{shin2016photon}, Rapp \cite{rapp2017few}, Non-local net \cite{peng2020photon}, U-net++ \cite{zang2021non} and STIN (ours) are shown in \tref{Tab:results}. An additional metric named the accuracy at a given threshold $\delta$ is defined as
\begin{align}
	&\text { percentage of } \hat{D}, \nonumber \\ 
	\text { s.t. } \ \frac{1}{N_x N_y} &\sum_{i}^{N_x} \sum_{j}^{N_y} \max \left(\frac{d_{i,j}}{\hat{d}_{i,j}}, \frac{\hat{d}_{i,j}}{d_{i,j}}\right) < \delta.
\end{align}

\section{Observation model}
\label{App:Observation}
The detected photon counts are modelled as a linear mixture of signal and background photons corrupted by Poisson noise \cite{shin2016photon}. Given the laser pulse waveform $s(t)$, the Poisson-process rate function for the ${(i, j)}$-th pixel of the object to be detected is
\begin{equation}
	r_{i, j}(t)=\eta a_{i, j} s(t-\frac{2z_{i, j}}{c})+n_{i, j},
	\label{eq:rt}
\end{equation}
where $n_{i, j}$ and $\eta \in(0,1]$ are the average count of the noise and the detection efficiency of the SPAD detector, respectively. Here, $a_{i, j}$ and $z_{i, j}$ are the albedo and the ground-truth depth of the $(i, j)$-th scanned point, respectively. Assume that the size of time bin is $\Delta$, the total number of time bins is $T$ and the number of illuminations is $n_{t}$, the statistical distribution for each time bin is formulated as
\begin{equation}
	h_{i, j,t} \sim \operatorname{Possion} (n_{t} \int_{(t-1) \Delta}^{t \Delta} r_{i, j}(x) \mathrm{d} x), \ t = 1, \cdots, T.
	\label{eq:poisson}
\end{equation}
The histograms of pixels are concatenated together to form a photon counting cube of a size $[T, N_x, N_y]$, where $T$, $N_x$, $N_y$ are the number of time bins, the horizontal and vertical resolutions, respectively.

\bibliography{sample}

\end{document}